\documentclass[oamsmath,amssymb,nofootinbib,twocolumn,preprintnumbers]
{revtex4}

\usepackage{epsfig}
\usepackage{amssymb}
\usepackage{bm}
\usepackage{hyperref}
\usepackage{color}
\usepackage{amsmath}

\newcommand{\be}{\begin{equation}}
\newcommand{\ee}{\end{equation}}
\newcommand{\bea}{\begin{eqnarray}}
\newcommand{\eea}{\end{eqnarray}}

\newcommand{\degree}{^\circ}

\newcommand{\fr}[2]{{\hbox{$ #1 \over #2 $}}}

\begin{document}

\title{Neutrino Signals in IceCube from Weak Production of Top and Charm Quarks}

\author{Vernon Barger$^1$}
\author{Edward Basso$^1$}
\author{Yu Gao$^{2,3}$}
\author{Wai-Yee Keung$^4$}
\affiliation{
$^1$Department of Physics, University of Wisconsin, Madison, WI 53706, USA\\
$^2$Mitchell Institute for Fundamental Physics and Astronomy, 
Texas A\&M University, College Station, TX 77843-4242, USA \\
$^3$Department of Physics and Astronomy, Wayne State University, 
Detroit, MI, 48201 USA \\
$^4$Physics Department, University of Illinois at Chicago, IL 60607\\
}

\begin{abstract}
Deep inelastic scattering of very high-energy neutrinos can
potentially be enhanced by the production of a single
top quark or charm quark via the interaction of a virtual $W$-boson exchange with a $b$-quark or $s$-quark parton in the
nucleon. 
The single top contribution shows a sharp rise at neutrino energies above 0.5 PeV and gives a
cross-section contribution of order 5 percent at 10 PeV, while single charm has a low energy threshold and contributes about 25 percent. Semileptonic decays of top and charm give dimuon events whose kinematic characteristics are shown.  The angular separation of the dimuons from heavy quark production in the IceCube detector can reach up to one degree.  Top quark production has a unique, but rare, three muon signal.

\end{abstract}

\maketitle

\section{Introduction}
The ultra-high energy cross-section for neutrino deep inelastic
scattering (DIS) has long been of theoretical interest. See,
e.g.\cite{DIS_list}.  The DIS
cross-section contributions due to the $b$-quark to $t$-quark transition and the $s$-quark to $c$-quark transition, mediated by
$W$-boson exchange\cite{bt_list}, may be observable in the IceCube experiment.  With the recently improved  determinations of the
$b$-quark and the $s$-quark parton distributions function (PDFs), single top-quark and single charm-quark 
production by neutrinos can be calculated with a high degree of
confidence and this is one objective of our study.

The IceCube experiment has recently reported results from 7 years
of data\cite{Aartsen:2016syr}.
Neutrino events with energies in the
range 240 TeV to 10 PeV are found
at a level that significantly exceeds the
atmospheric neutrino background which is steeply falling with increasing energy. These observations have sparked interest
in possible origins of the high energy events \cite{Anchordoqui:2014etl}, including astrophysics sources, such as AGN and star-burst galaxies \cite{Aartsen:2016ngq}, or
new physics, such as leptoquarks \cite{Barger:2013pla,Dutta:2015dka,Dey:2015eaa}
and the decays of very-long-lived neutral particles associated
with quasi-stable dark matter\cite{Anchordoqui:2015lqa, Feldstein:2013kka, Bai:2016bls, Ahlers:2016abb}.

The Standard Model contributions from the production of
a single top-quark and a single charm-quark  will enhance the 
DIS neutrino cross at PeV-energies and thus these contributions are relevant to IceCube observations.  We evaluate their DIS contributions and consider the characteristics of dimuon events associated with the semileptonic decays of the t-quark and c-quark.

We begin with a brief overview of the IceCube experiment and datasets.  IceCube is a 1 km$^3$ photomultiplier-instrumented detector located in the South Pole ice sheet.  The detector measures the total Cherenkov light emission in a high-energy neutrino event. The produced leptons and hadrons contribute to the observed Cherenkov light. The IceTop array of ice tanks on the surface is used to detect and reconstruct air showers; it thereby vetoes the large cosmic muon backgrounds.

There are two classes of events:                                                                                                               

1)``Track-like" events are those with a highly energetic muon produced in the interaction of a $\nu_\mu$ within the detector or in the surrounding ice or rock. In addition to rejection of cosmic muon backgrounds by IceTop, the Earth also serves as a filter to eliminate cosmic muon backgrounds.  Muons with arrival directions above 85 degrees in zenith angle must originate from neutrino interactions, even if the muon track originates outside the detector volume. 

 2) ``Shower-like" events are those with an electromagnetic shower that is contained in the detector
 but without a muon track. These events are due to $\nu_e$ or $\nu_\tau$ charged-current events , as well as neutral current events. 
 
 The Class 1 track events are up-going in the detector.  They are essentially free of the atmospheric background, but they provide only partial sky coverage. There is a significant loss of the very high energy neutrino flux in the propagation of the neutrinos through the Earth. 
 
The Class 2 shower events are required to have the visible electromagnetic energy confined within the detector volume.  The cosmic muon background is rejected by IceTop.  The Class 2 events have full sky coverage.

The contributions to the atmospheric neutrino flux from pair-production of
charm particles by the strong interaction have been considered \cite{Bhattacharya:2016jce, Halzen:2016thi, Gauld:2015yia, Laha:2016dri}, with the conclusion that this source cannot explain the excess of events observed by IceCube above 30 TeV\cite{Halzen:2016thi}.

In the Class 1 track events, the most probable neutrino
energy cannot be precisely determined because the high-energy muon often passes through
and exits the detector. However, the neutrino direction of the track events
is well determined to less than 0.5 degree. 

In the Class 2 shower events, the energy of the incident neutrino is reasonably well determined, while the neutrino direction has large uncertainty (with a median uncertainty of 10 degrees). 

Thus, the two Classes of events are complementary in their physics information.  The neutrino flux is steeply falling up to 100 TeV, as expected for neutrinos of atmospheric origin. Above 240 TeV, the neutrino flux has a flatter energy spectrum that is consistent with a $E_\nu^{-2}$ power law,
typical of an astrophysics Fermi acceleration mechanism of cosmic
rays~\cite{Waxman:1998yy}.  Whether there is a maximum energy cut-off of the neutrino flux remains an open question.

The three most energetic shower events have energies of
1.041 PeV, 1.141 PeV and 2.0 PeV, with 15\% energy resolution.
A track event was found with an exceptionally high-energy muon and 
$2.6 \pm 0.3$ PeV deposited energy.
These are the highest energy neutrinos ever recorded by any experiment.
The high-energy neutrino flux inferred by IceCube depends on the effective area of the detector, under the assumption that the neutrino inclusive cross-section can be accurately modeled by charged-current and neutral-current DIS on light-quark flavors. Our study evaluates the impact of the $b$-quark to $t$-quark and the $s$-quark to $c$-quark transitions, treating the
$b$-quark as a massless parton in the proton~\cite{CTEQ:2016bdf,Olness:2016cgo} in the 5-flavor formalism.  In addition, we simulate the muon distributions in dimuon events for a further probe of heavy quark contributions.  Our focus is on events in which the deep inelastic interaction on a proton target of a $\nu_\mu$ gives a fast primary muon.

\section{Slow scaling in top-quark production}

In a 4-flavor parton scheme (4FS),
the leading-order (LO) partonic process for the QCD production of a $b$-quark
is  $gluon$ to $b \bar b$ and the top-quark is produced from the b-quark in an overall 2 to 3 particle process.
In the 4FS, the integration over the final-state bottom-quark momenta
leads to logarithmic dependence on $m_b$. In a 5-flavor scheme (5FS),
these logarithms are re-summed to all orders in the strong coupling
into a $b$-quark parton distribution function (PDF). 

In the 5FS, the b-quark mass is set to zero, and all collinear
divergences are absorbed into the PDF through mass factorization. The dependence on the b-quark mass is encoded as a boundary condition on the Renormalization Group Equations. In the 5FS, top-production in DIS is a 2 to 2 particle process.  We
adopt the 5FS for our calculations for effectiveness, since either the 4FS or 5FS scheme should give the same cross-section.~\cite{Kilgore:2003hk}
A similar use of the $b$-parton PDF in the calculation of Higgs production at colliders can be found in
~\cite{Kilgore:2003hk,Maltoni:2003pn}.

The leading order Feynman diagram for top-quark production in the 5FS via the
weak charged-current neutrino interaction is shown in
Fig.~1, along with the top-quark decay to a b-quark and a real W-boson.
\begin{figure}

\includegraphics[scale=0.7]{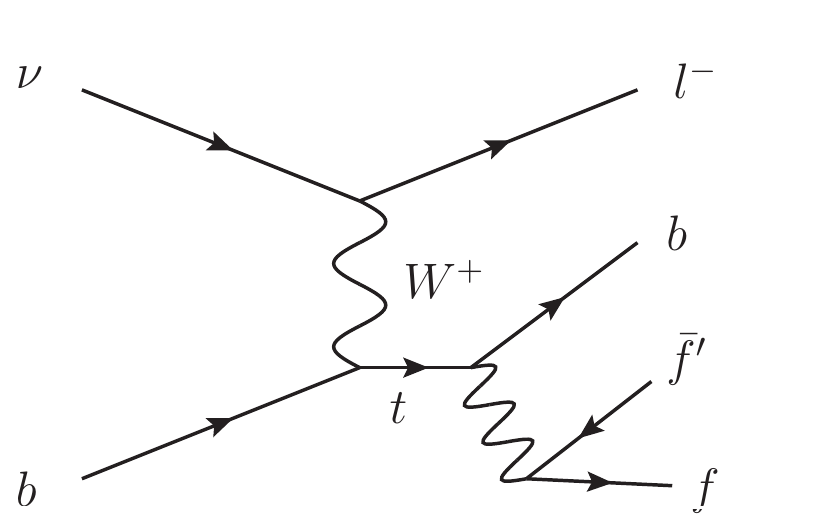}
\caption{
Leading order Feynman diagram for neutrino production
of the $t$-quark from the $b$-quark parton in the nucleon,
$\nu b \to \ell t$.
The W-boson decays to a fermion and an anti-fermion}
\label{fig:feynman_diagram}
\end{figure}

The charged current subprocess $\nu b \to \ell t$ gives the deep
inelastic $t$-quark production cross-section. 
In the excellent approximation that the quark mixing matrix element $V_{tb}=1$, the differential DIS cross section is given by
\begin{equation} 
\frac{d\sigma}{dx dy}=
\frac{G_F^2 (\hat s-m_t^2) m_W^4}{\pi (Q^2+m_W^2)^2} b(x',\mu^2) \ ,
\end{equation}
where the momentum transfer $q=p_\nu-p_\ell$ sets the scale
$Q^2=-q^2>0$. The Bjorken scaling variables are 
$x=Q^2/2 p\cdot q$ and $y=p_N\cdot q/m_N$, with
$Q^2 = sxy$;
 $y=(E_\nu-E_\ell)/E_\nu = E_h/E_\nu$ is the fraction of the neutrino energy that is transferred to hadrons. The CM energy squared of $\nu N$ scattering is $s=2m_NE_\nu$, neglecting the small $m_N^2$ contribution.
From kinematics, the fractional momentum of the $b$-parton is
$x'=x+{m_t^2/ys}$.
The subprocess CM energy squared is $\hat s=(p_\nu+p_b)^2=x' s$.
The domains of the $x,y$ variables are
\begin{equation}
m_t^2/s < y < 1 \  {\ \rm and \ } 0<x<1-\fr{m_t^2}{sy} \ .
\end{equation}
Note that $b(x',\mu^2)$
is evaluated at the slow scaling variable,
i.e. $x'$.

After variable substitutions, we also obtain the formula
\begin{equation}
\frac{d\sigma}{dx dy}=
\frac{G_F^2 (2 m_N E_\nu x + m_t^2/y -m_t^2) m_W^4}
{\pi (m_W^2+2 m_N E_\nu xy)^2}
b(x+\fr{m_t^2}{sy},\mu^2) \ , \end{equation}
with $y(1-x)>m_t^2/s$. Note that the numerator factor
$(xs + m_t^2/y -m_t^2) \longrightarrow xs$
when $\hat s \gg m_t^2$, and thus
$xb(x)$ is obtained in Eq. (3) well above threshold.
A similar formula applies to the anti-neutrino case.
In our calculations we take $m_t$ for both factorization and renormalization scales, as found in other applications to reproduce NLO and NNLO results in a LO calulation\cite{Kilgore:2003hk,Maltoni:2003pn} 

\medskip
\section{Cross sections and $y$-distributions}
\label{sect:Cross sections and y-distributions}

The calculated neutrino DIS charged-current cross sections are shown in Fig.~2
versus the neutrino energy. The upper curve is the result for 4 light parton flavors $(u,d,s,c)$; NLO QCD corrections\cite{Dobrescu:2003de} are found to be -1\% of the LO result at all energies and thus are insignificant.  However, the calculated DIS cross sections are subject to possible overall uncertainties associated with the PDFs, but again these will be independent of neutrino energy. The lower curve  in Fig.~2 is the contribution from top-quark production. Above 10 PeV the top-quark cross section approaches 5 percent of the usual CC result.  The middle curve is the contribution from of c-quark production from the s-quark.  Single charm production is about 25 percent of the total DIS.  The weak production of the charm quark from the strange quark in the proton has a low neutrino energy threshold and the energy dependence is quite unlike the steep rise with energy of top quark production.

\begin{figure}
\includegraphics[scale=0.4]{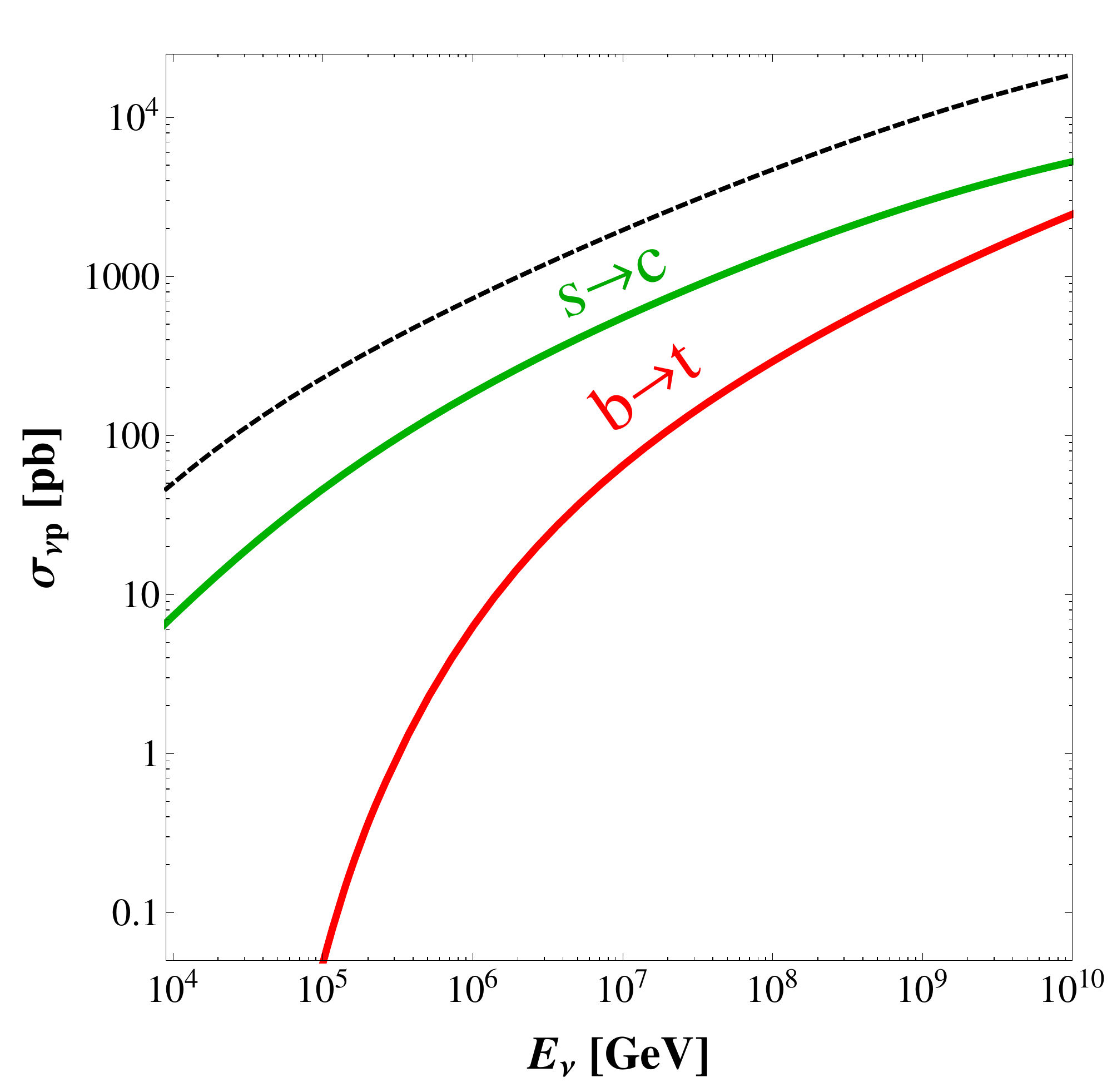}
\caption{Deep-inelastic $\nu_\mu$ cross-section for charged-current scattering on a proton target. The upper curve is the standard result for $u, d, s, c$ partons. The middle curve shows the cross-section from the $s$-quark to $c$-quark process. The bottom curve is the DIS contribution from scattering on the $b$-quark parton to produce the $t$-quark reaction. 
}
\label{fig:total_cross_section}
\end{figure}
Physics with a high threshold energy, like the top, will first
become evident at low $x$ and high $y$. The distributions in the scaling
variable $y = 1 - E_\mu/E_\nu$ are shown in Fig. 3a, for three choices
of neutrino energy:
0.1 PeV (close to the threshold for top production), 1 PeV (an energy for which the background from atmospheric neutrinos is negligible) and 10 PeV (where the $y$-distribution for top production approaches the shape of the usual result of 4-quark flavors). The $y$-distribution at 0.1 PeV clearly exhibits the kinematic suppression from the top-quark threshold. 

The theoretical  distribution in the $y$ variable from scattering on light partons has been used by the IceCube collaboration in estimating the neutrino energy of through-going muon events from the Cherenkov light. Figure 3b compares the average-$y$ values, $\langle y\rangle$, for production from 4-quark flavors with that from top-production. There are substantial differences in $\langle y\rangle$ for neutrino energies of 1-10 PeV. Thus, since $E_\nu$ = $E_{\rm hadron}/y,$ a  higher neutrino energy would be inferred for an event assuming production from light partons then would be the case  if it is a top-quark event. However, the importance of this effect should be modest, since the top cross section at a neutrino energy of 1 PeV is only at the 5 percent level.  At the highest energies in Fig. 3b, the 4-flavor and $t$-quark results for 
$\langle y \rangle$ are converging, since sea quarks then dominate the cross
sections. We note that the trend towards smaller $y$ with increasing
energy, for both the usual CC and $t$-quark cross-sections, is a
consequence of the the $Q^2$ dependence of the $W$-propagator, which
suppresses high-$y$ contributions.

\begin{figure}
\includegraphics[scale=0.4]{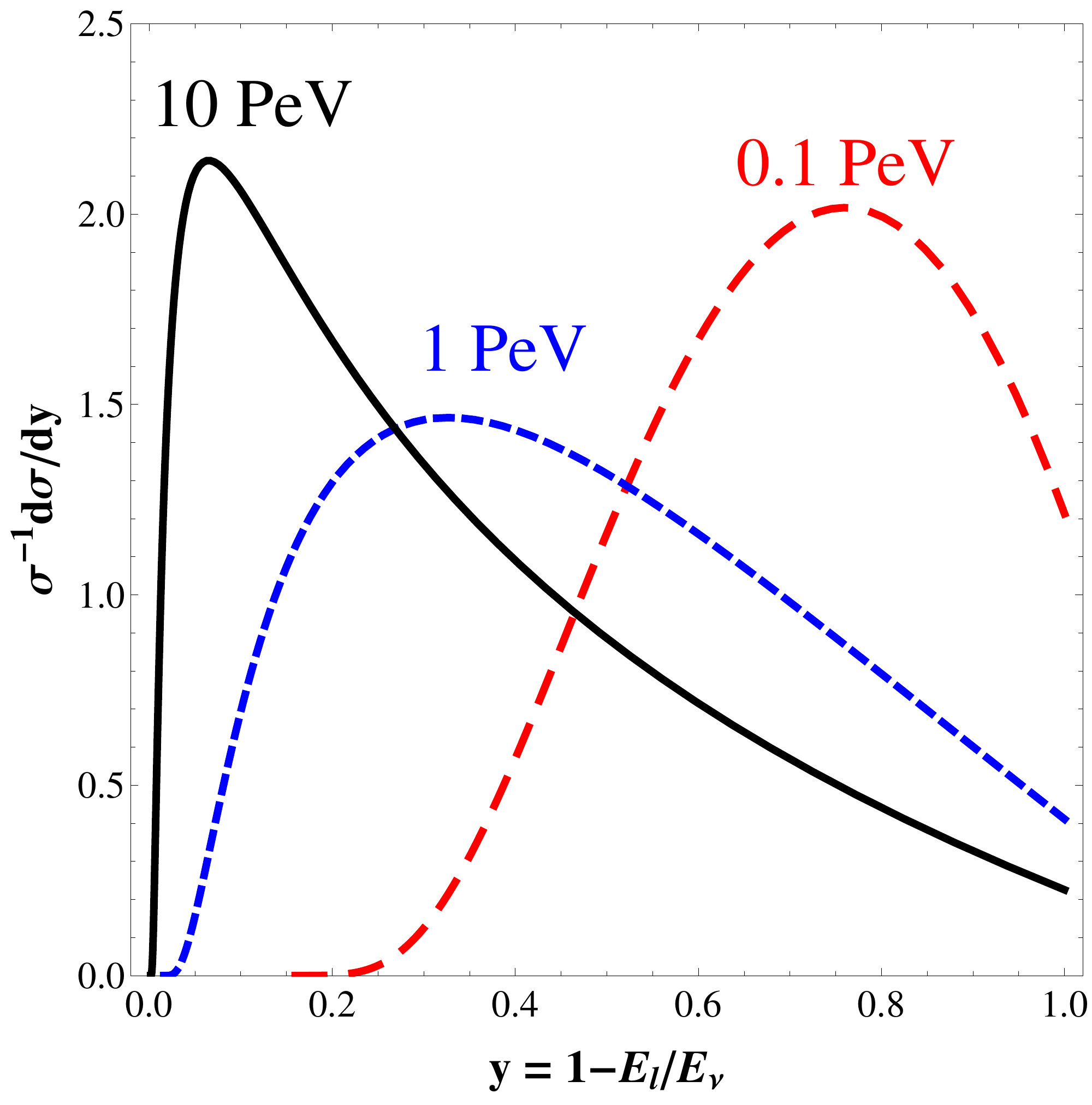}
\includegraphics[scale=0.4]{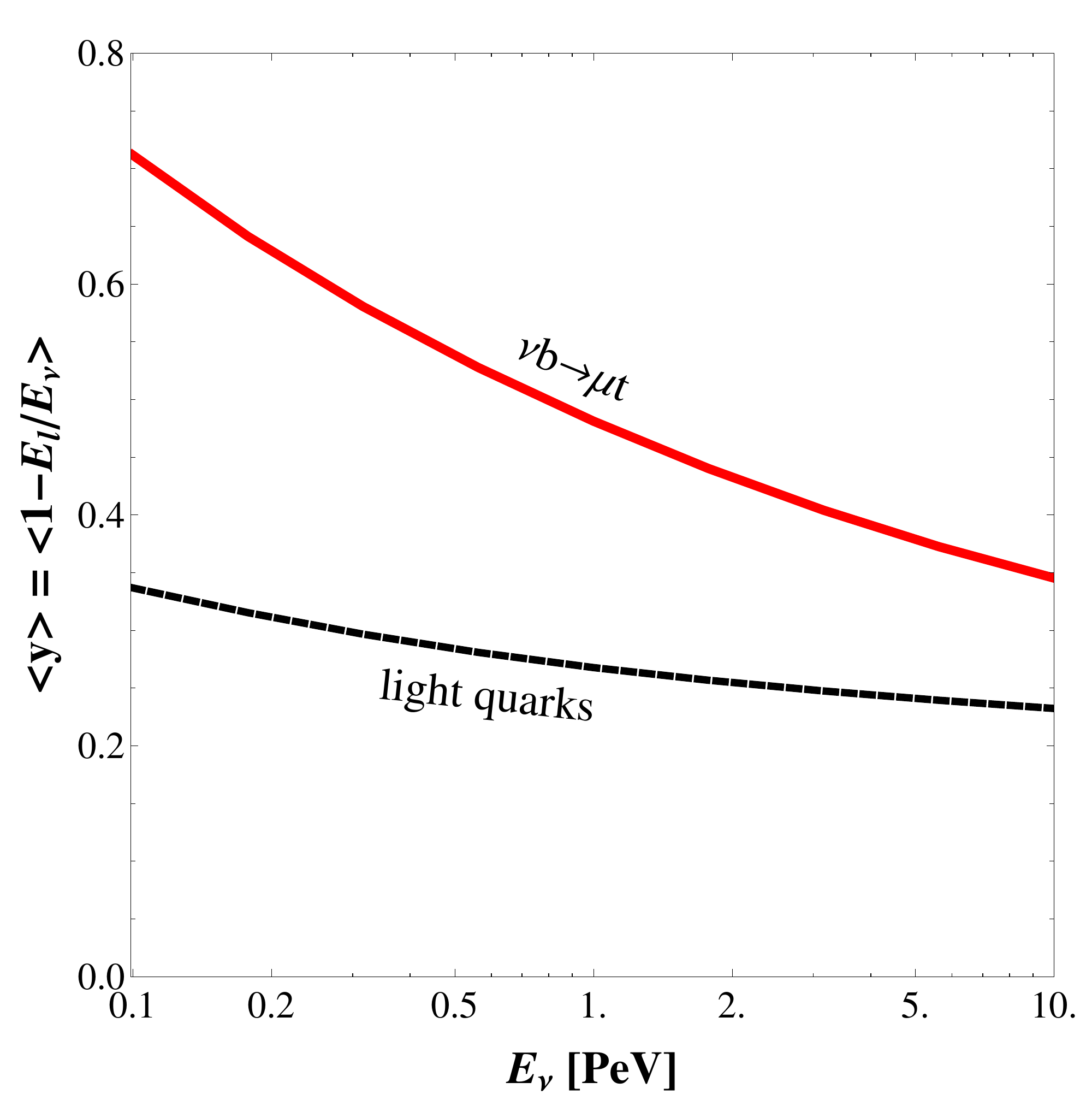}
\caption{
(a) Distributions versus the scaling variable $y = 1 - E_\ell/E_\nu$ in
charged-current neutrino deep-inelastic scattering, at neutrino
energies of 0.1, 1, and 10 PeV; the distributions are normalized to unity to facilitate
comparison of the shapes; (b) Average $y$ versus neutrino energy for
scattering on $u, d, s$ and $c$ (dashed curve) and for scattering on the
$b$-quark parton to produce the $t$-quark (solid curve).}
\label{fig:invariables}
\end{figure}
\medskip

\section{Dimuon events}
\label{sect:Dimuon events}

In addition to a primary muon in the DIS of $\nu_\mu$s, the decays of a top quark into $B$-mesons  or a charm quark into $D$-mesons will lead to additional muons in about 10 percent of heavy quark events.  In the following we label the most energetic muon in an event as $\mu_1$, which mostly will be the primary muon from the neutrino production vertex and that of the second most energetic muon as $\mu_2$, which will mostly be the muon from the decay of the heavy quark.

 At high neutrino energies, $\mu_2$ will typically also have moderately high energy due to the large Lorentz boost from the center-of-mass frame to the laboratory frame.  We simulate the predicted kinematic distributions of these muons from heavy quark decays using MadGraph5~\cite{Alwall:2014hca} 
 for the production cross sections and PYTHIA6~\cite{PYTHIA6} for the hadronizations into $B$ and $D$ mesons as well as their decays.  Top quarks decay before hadronization, so we include the spin correlations of production and decay in that case. 
 
\begin{figure}[h]
\includegraphics[scale=0.6]{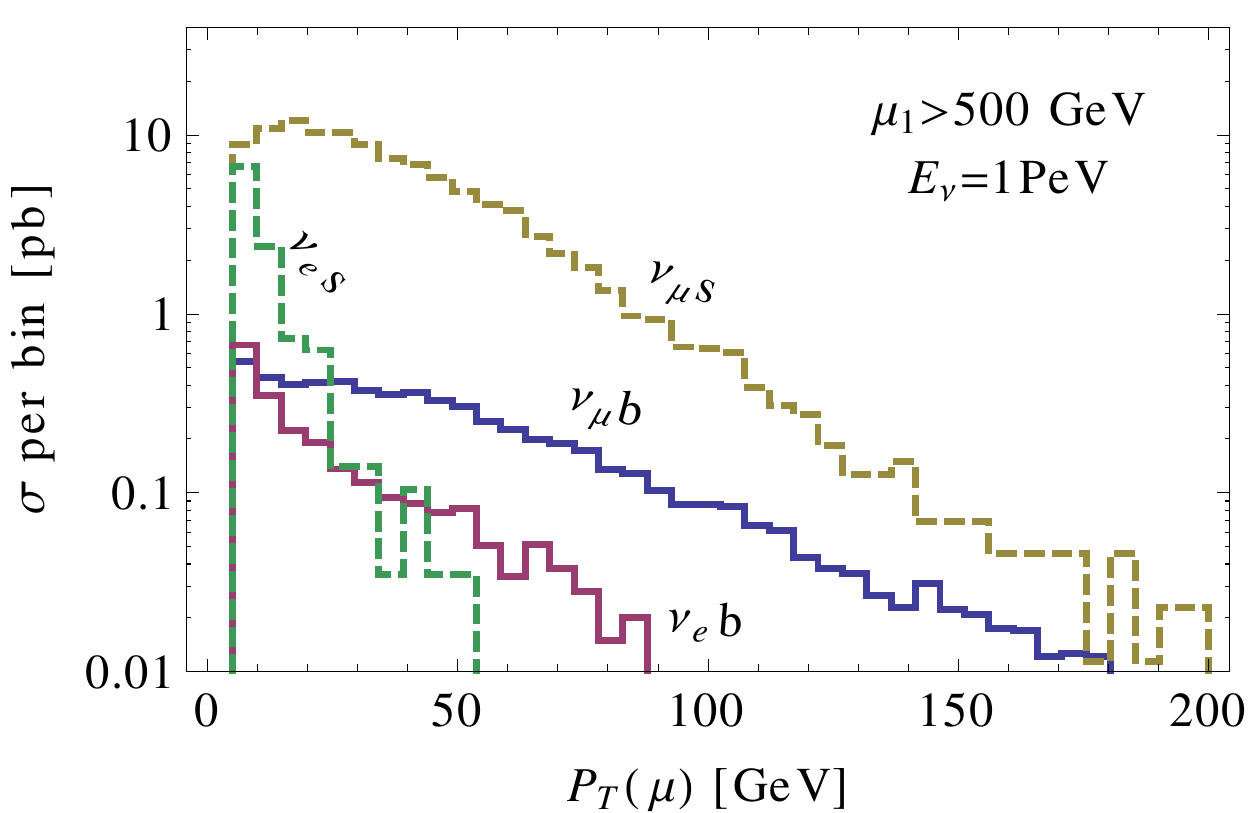}
\includegraphics[scale=0.6]{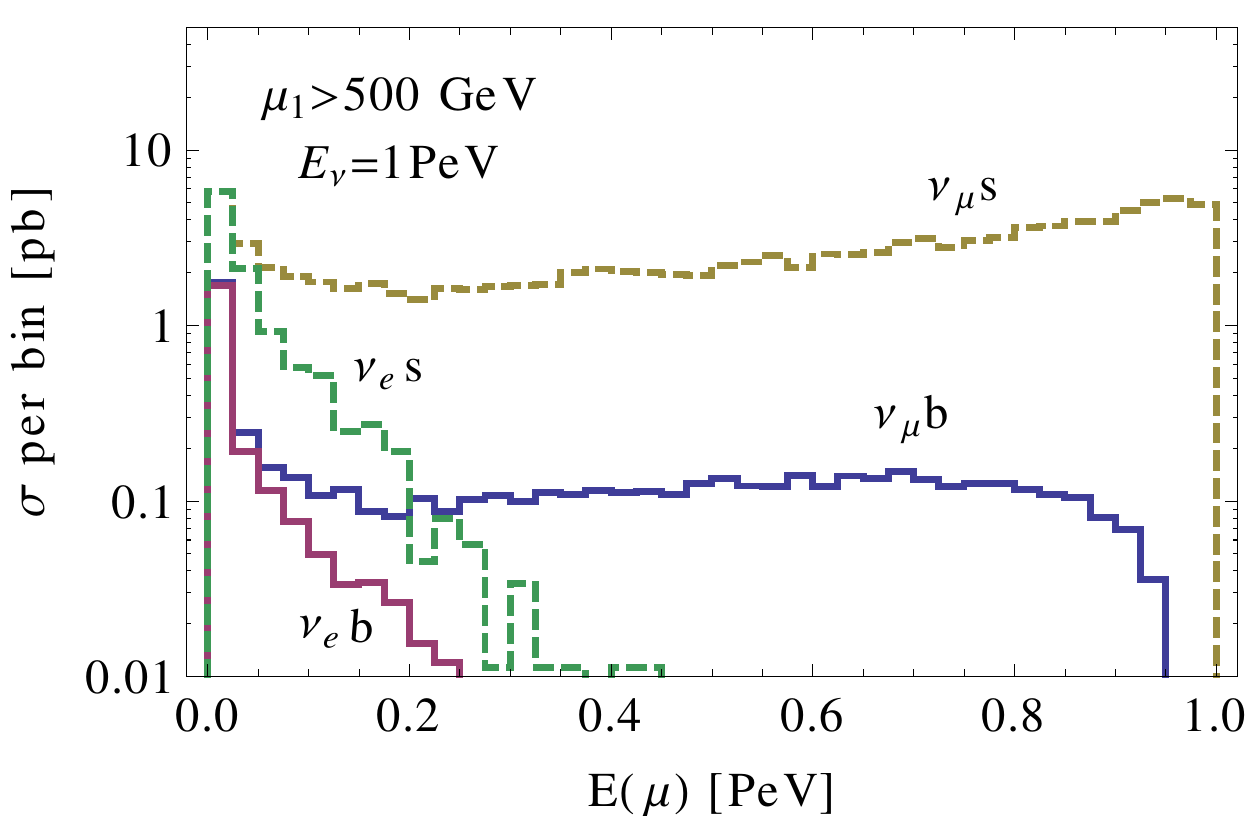}
\caption{Muon $P_T$ and energy distributions for incoming neutrino energy at 1 PeV. The labels of the curves denote the type of neutrino and the target sea-type quark. The highest energy muon in an event is required to have energy greater than 0.5 PeV.}
\label{fig:muon_pt_ene}
\end{figure}

\begin{figure}[h]
\includegraphics[scale=0.6]{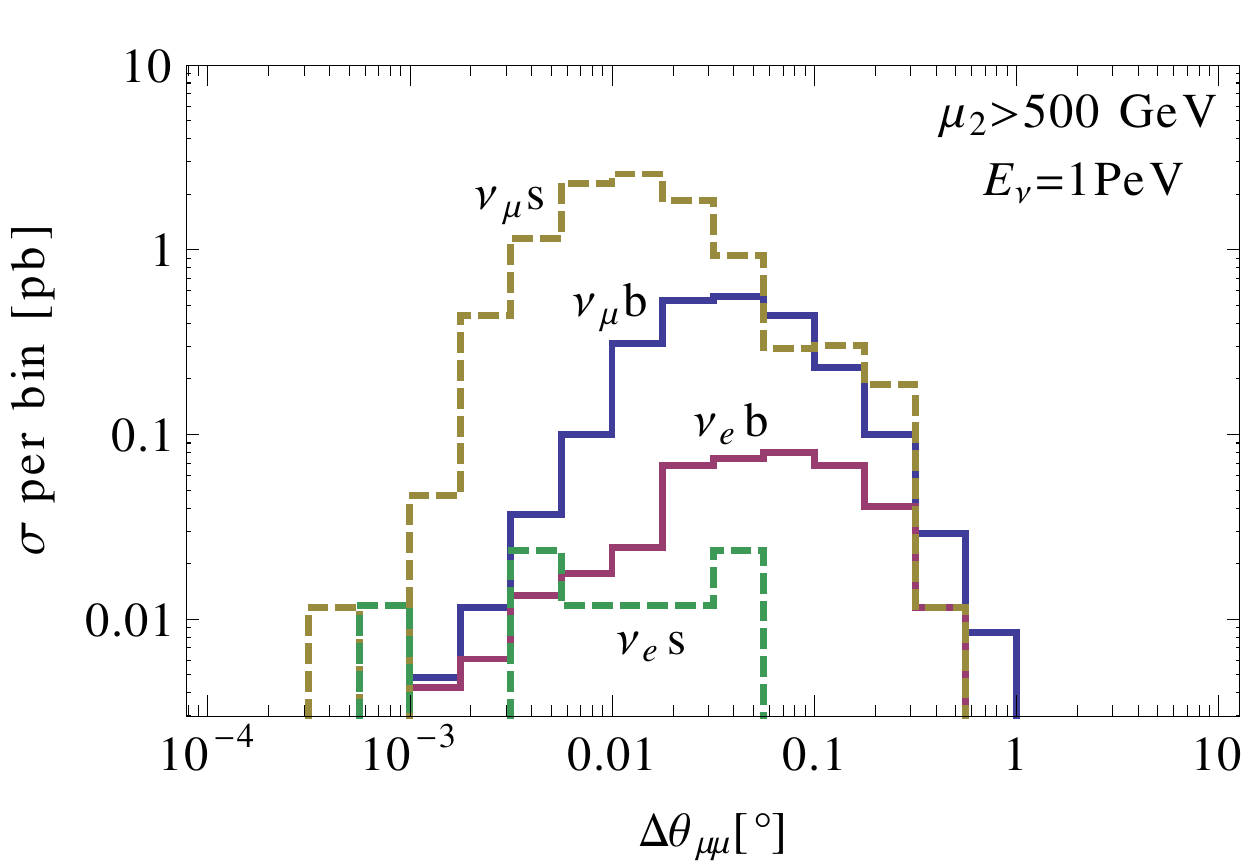}
\caption{Angular separation between the leading dimuons. A minimum energy requirement of 500 GeV is imposed on the second most energetic muon ($\mu_2$).  Each curve is normalized to the fraction of the scattering cross-section that yields at least two muons with a $\mu_2$ energy above 500 GeV.
Colors and labels are the same as in Fig.~\ref{fig:muon_pt_ene}
}
\label{fig:angular_separations}
\end{figure}

The muon transverse momentum and energy distributions are shown in Fig.~\ref{fig:muon_pt_ene} and the angular separations of the two leading (in energy) muons are shown in Fig.~\ref{fig:angular_separations}, at an incoming neutrino energy of 1 PeV. In each figure,  $\nu b$ represents a $b$ to $t$ conversion, $\nu s$ represents $s$ to $c$ conversion, etc.   All muons in the final state, both from the neutrino-vertex and those from a real $W$-boson, when present, as well as the $B, D$ decays, are included. The muons from decays of the longer lived pions and kaons are not included as they will lose energy quickly and range out during their propagation in the ice or rock.

The radiation of a W or Z boson, from internal and external particles of the lowest order weak processes, are also a potential source of multi-lepton events when the W or Z decay to muons (or the W and Z decay to c and b quarks that subsequently decay to muons). We have calculated these contributions to dimuon events and found that they are about an order of magnitude smaller than dimuons from top production, at neutrino energies above 1 PeV; these contributions are about equal to the top contribution at a neutrino energy of 0.3 PeV, due to the kinematic suppression of top production near its threshold. 

The corresponding distributions from $\nu_e$ DIS are shown in Fig.~\ref{fig:muon_pt_ene} as a useful comparison.  $\mu_1$ of an $\nu_e$ event has the energy distribution of  $\mu_2$ in a $\nu_\mu$ event.  The energy of the fastest muon is required to exceed 0.5 PeV in these plots.  For $\nu_e$ DIS, 46 percent of b to t events and 11 percent of s to c events satisfy this energy requirement.  

The observation of an energetic $\mu_2$ will signal a heavy quark event.  The charm contribution dominates over top by about a factor of 10 for a 1 PeV primary neutrino energy, so disentangling the top signal from energy distributions alone would be challenging, but the angular separation of the two muons is more favorable, as discussed below. The proposed Generation 2 expansion of the IceCube detector will provide  a factor of 10 increase in events, along with sensitivity to higher neutrino energies, which may make a partial distinction possible of the top and charm dimuon signals.

Due to the very small deflection angle of a primary muon from the neutrino direction, the $P_T$ 
of $\mu_2$ with respect to the neutrino is essentially the same as relative to the $\mu_1$ direction.  
In $\nu_e$ events, both muons originate from the hadron vertex and consequently the $P_T$ distribution is much softer, as can be seen in Fig.~\ref{fig:muon_pt_ene} .

\begin{figure}[h]
\includegraphics[scale=0.6]{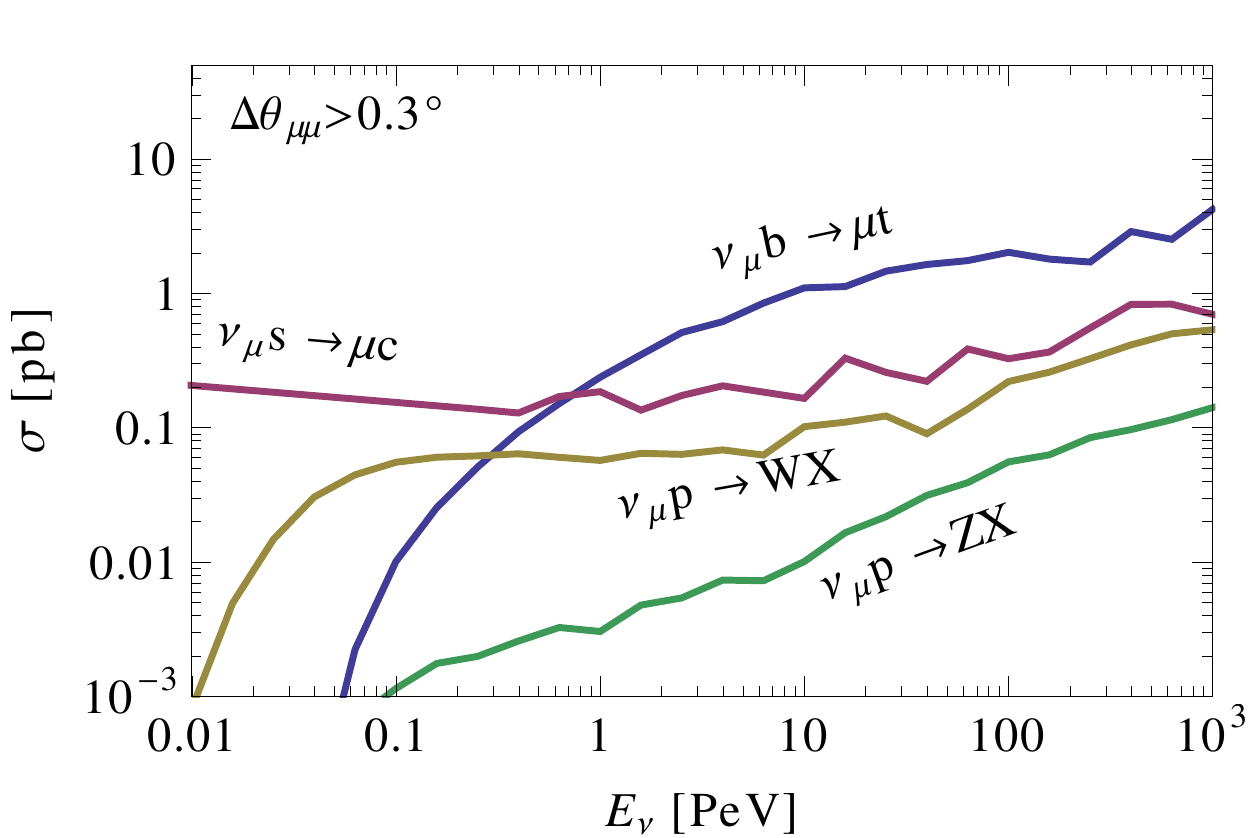}
\caption{Cross-section of channels that produce muon pairs with angular separation greater than 0.3$~\degree$. Muon energies are required to be greater than the IceCube threshold of 70 GeV. The inclusive branchings of $W, Z$ into muons are included in each channel.}
\label{fig:xsec_after_ang_cut}
\end{figure}

The angular separation between $\mu_1$ and $\mu_2$ is typically small, due to the high Lorentz boosts, yet some events will have an angular separation as large as  $1$ $\deg$, as shown in Fig~\ref{fig:angular_separations}.  The $\mu_2$ from top production is more likely to lead to a larger angular di-muon separation than is the case for charm. 

When the two muons originate in the rock or ice prior to reaching the detector, the spatial separation of their tracks within the detection may be resolved.    
IceCube can distinguish the tracks of two muons when their opening angle is greater than 0.28 deg (0.005 rad): see Fig.~2 of Ref.~\cite{Aartsen:2016syr}. With an angular separation cut $>$ 0.3 deg imposed, the dimuon cross-section from charm is about 0.2 pb, for neutrino energies between 10 TeV and 1 PeV, 
as shown in Fig.~6.  
For 2 muons that are separated by less than 0.3 deg, the energy inferred from the emitted light will exceed that from the primary muon.  
A full detector simulation is necessary to properly judge the ability of IceCube to distinguish the tracks of the two muons.

Also, trimuon events can arise from $\nu_\mu$ production of the top quark and its decays to $W$ plus a $b$-quark (with the probability increasing from
9\%  at $E_\nu\sim$PeV  to 12\% in the high $E_\nu$ limit), with a primary muon from the neutrino vertex, a muon from leptonic $W$-decay and the third muon from $B$-decay; such an event is rare (of order 10$^{-2}$) but nearly background free. In trimuon events, the higher order electroweak contributions are about an order magnitude below the trimuon contribution from top production, for neutrino energies above 1 PeV.  Serendipitous discovery of the trilepton signature of top production is possible if close-by muon tracks can be distinguished.

\medskip

\section{Summary}
\label{sect:Summary}

Weak production of the top-quark gives an increase of order 5 percent in the neutrino deep inelastic scattering cross section at PeV energies, with a sharp threshold rise.  Single charm production contributes about 25 percent. With semileptonic decays, both give rise to di-muon events with distinctive kinematic characteristics.  The tracks of the two muons may be separated by up to one degree in the IceCube detector.  The top quark gives a unique and background free, but rare, three muon signal.  The discovery of energetic multi-muon events by IceCube will be a physics tour de force of great interest.

\medskip
\section{Acknowledgements}

VB thanks Fred Olness for illuminating discussions about the $b$-quark PDF and Francis Halzen for discussions about IceCube capabilities.
This work is partially supported by the U.S. Department of Energy
under grants DE-AC02-06CH11357 and DE-FG-02-12ER41811.
YG thanks the Mitchell Institute for Fundamental Physics and Astronomy (MIFPA)
and Wayne State University for support.

\newpage

\end{document}